# Peak effect in YBCO crystals: Statics and dynamics of the vortex lattice

G. Pasquini and V. Bekeris
*Laboratorio de Bajas Temperaturas, Departamento de Física,*
*FCEN, Universidad de Buenos Aires; CONICET. Argentina.*

## Abstract

Oscillatory dynamics and quasi-static Campbell regime of the vortex lattice (VL) in twinned $YBa_2Cu_3O_7$ single crystals has been explored at low fields near the peak effect (PE) region by linear and non-linear *ac* susceptibility measurements. We show evidence that the PE is a dynamic anomaly observed in the non-linear response, and is absent in the Labusch constant derived from the linear Campbell regime. Static properties play a major role however, and we identify two $H(T)$ lines defining the onset and the end of the effect. At $H_1(T)$ a sudden increase in the curvature of the pinning potential wells with field coincides with the PE onset. At a higher field, $H_2(T)$, a sudden increase in linear *ac* losses, where dissipative forces overcome pinning forces, marks the end of Campbell regime and, simultaneously, the end of the PE anomaly. Vortex dynamics was probed in frequency dependent measurements, and we find that in the PE region, vortex dynamics goes beyond the description of a power law with a finite creep exponent for the constitutive relation.

## I. INTRODUCTION

The Peak Effect (PE) in the critical current density ($J_c$) in both low and high temperature superconductors has been subject of a large amount of experimental and theoretical work in the last decade. An anomalous increase is observed above an onset field $B_{on}$ or temperature $T_{on}$, until $J_c$ reaches a maximum at $B_p$ (or $T_p$) above which a fast decrease in the measured critical current density occurs, just before reaching either the melting or the upper critical field line.

The origin and nature of this phenomenon are still controversial issues. A unified phenomenological picture based in an order-disorder (O-D) transition for the vortex lattice (VL) from a quasi-ordered Bragg glass (BG) to a disordered phase with increasing topological defects, using the Lindemann criterion has been developed[1], and explains a broad amount of experimental results. However, the underlying physics and the nature of this transition is still controversial. Since the first description by Larkin and Ovchinnikov[2], several theoretical pictures have been proposed: The crucial role of dislocations and its consequence in VL dynamics in the PE region was highlighted[3–5]. Recently, a scenario where VL shows no order-disorder phase transition but a continuous transformation to an amorphous state has been proposed[6]: in this picture the PE arises from a change in the dynamics (possible jamming) of vortex matter. Other authors[7] claim that a change in the pinning regime is the origin of the PE. Probably, the large collection of proposed models are a consequence of the different nature of the PE in the various materials, as is becoming clear from the increasing amount of experimental evidence.

In traditional superconductors, neutron diffraction experiments show a clear change in the structure factor of the VL indicating an O-D transition[8]; the thermal hysteresis observed in this transition should indicate a first order character. In $NbSe_2$, experiments of magnetization assisted by a shaking *ac* field[9] as well as transport measurements in the Corbino geometry[10] suggest a first order transition. Surprisingly, experiments of Bitter decoration[11] show that the disordered VL is not an amorphous (or a pinned liquid) but a polycrystal. Several works indicate a crucial role of the dynamics of the VL in the PE in this material: a change from an elastic to a dislocation mediated plastic regime[12] and a collective to individual pinning transition[13] at the onset of the PE have been proposed.

In high temperature superconductors (HTS´s), the results are still more controversial. In BSCCO samples, the existence of a first order transition at the PE has been established by experiments of differential magneto-optic (MO)[14], and supported by local magnetization measurements[15]. However, the crucial role of dynamics and creep in the PE, evidenced by the time dependent results has been known for a long time[16]. The dramatic decrease in the measured $J_c$ before the onset of the PE should be mainly due to creep effects, but it seems to be a precursor to a genuine change in the elastic properties of the VL that reduces the vortex correlation length causing the PE[17]. Recently, direct observation by MO imaging of the magnetic induction profile in the vicinity of the PE, shows a larger $J_c$ in the outer region of the samples, that some authors ascribe to a transient metastable disordered vortex phase penetrating through the sample surface[18], and others to a coexistence between two bulk phases in a local first order transition[19].

In non-layered HTS´s, a direct observation of a VL transition from a BG to a short correlated vortex glass phase (VG) at $B_{on}$ has been recently reported in the LSCO system by muon spin resonance[20].

In the case of YBCO crystals, the first observation of the PE was made more than 10 years ago[21]. However, at the present time the role of disorder and twins in the occurrence of the PE is controversial, and not completely understood[22,23], and fundamental discussions describing a thermodynamic[26] or a dynamic picture are not settled. In ultrapure samples, an unusually giant and abrupt PE in the non-linear ac susceptibility has been reported[27].

Interestingly, VL history effects are concurrently observed in the PE region in a wide range of superconducting materials, and are probably closely related to the same kind of phenomena. In fact, in the vicinity of the PE, both the mobility of the VL and the measured $J_c$ are found to be dependent on the dynamical history of the sample in both low $T_c$[10,28] and high $T_c$[23,29,30] materials. In this framework, a great amount of work was devoted to understand if this phenomenology is dominated by surface or bulk pinning properties. In $NbSe_2$[10] and BSCCO[24,25] samples, surface and geometrical barriers seem to play a fundamental role. However, in materials with moderate anisotropy and high density of pinning centers, as is the case of well oxygenated twinned YBCO, transport properties at $H_{dc} >> H_{c1}$ are generally well described by bulk pinning forces.

In ac susceptibility experiments, the PE is observed as an anomalous increase in shielding capability or a decrease in losses, also usually interpreted as an anomalous decrease in VL mobility. Recent experiments in YBCO crystals, have shown that the mobility of the VL increases after assisting the system with a symmetric ac field (or current)[30] of moderate amplitude[31]. On the other hand, when vortices are assisted by an asymmetric ac field, the VL becomes less mobile[30]. This salient feature indicates that these effects cannot be ascribed to an equilibration process, but have their origin in the oscillatory character of vortex dynamics.

In our recent work, the solid VL was prepared with different dynamic histories and explored in the linear Campbell regime, using a very small ac field. We have shown that the dynamic history not only determines the degree of mobility, but also directly modifies the effective pinning potential wells, leading to different history dependent static VL configurations (VLCs)[32,33]. However, in samples where the PE is clearly displayed in the non-linear response, the anomaly (understood as the non-monotonous T dependence) is absent in the Campbell regime[23,33,34]. We think that this fact is a key point to understand the nature of the PE.

In this paper we present new results that will throw light onto this controversial subject. We explore the static and dynamic behavior of the vortex lattice in YBCO crystals with the dc magnetic field tilted out of the twin boundaries. By performing sensitive linear and non-linear ac susceptibility measurements we compare the behavior of VL mobility and the effective pinning potential wells in the region of the PE, at low fields and high temperatures. Assisted by large enough ac fields, vortices perform inter-valley motion and the non-linear penetration depth is directly related to the vortex mobility and the measured critical current density, while for the lower amplitudes, intra-valley motion dominates and the shape of the effective pinning potential well is probed. From our present results, the PE in YBCO crystals originates from a drastic change in the dynamics of the VL. This dynamic change, however, is correlated with a genuine modification in the dependence of the Labusch constant with magnetic field, probably due to an increase in the amount of dislocations.

The paper is organized as follows: In section II the experimental array and the numerical procedure followed to analyze the Campbell regime are described. Results and discussions are presented in section III and conclusions are drawn in section IV.

## II. METHODOLOGY

### A. EXPERIMENTAL ARRAY

The samples used were $YBa_2Cu_3O_7$ twinned single crystals[35] (typical dimensions $0.6 \times 0.6 \times 0.02 mm^3$) with $T_c \sim 92\ K$ at zero dc field and $\Delta T_c \sim 0.3\ K$. (10%-90% criterion). Results shown in this work correspond to a roughly cuandrangular shaped sample with a critical temperature (defined as the middle point of the linear ac susceptibility transition at 0 dc field) $T_c = 91\ K$. Similar studies have been performed in other two samples, and showed an alike phenomenology.

Global ac susceptibility measurements were carried out with the usual mutual inductance technique. The static magnetic field $H_{dc}$ is provided by a magnet that can be rotated relative to the sample, and was oriented out of all the groups of twin planes. The measuring ac field is parallel to the crystal c axis. All the experimental curves of χ' and χ" at each amplitude and frequency were normalized by the same factor, corresponding to a total step $\Delta \chi' = 1$ between the normal and superconducting response with $H_{dc} = 0$.

### B. NUMERICAL PROCEDURE

Pinning potentials were explored measuring the linear real penetration depth $\lambda_R$ in the Campbell regime[36], for small ac field amplitudes.

In a general case, the linear ac response is determined by the complex frequency dependent penetration depth $\lambda_{ac}(f) = \lambda_R + i\lambda_I$[37]. The function $\chi(\lambda_{ac})$ depends on the sample geometry. In the particular case of the Campbell regime vortices oscillate inside their effective pinning potential wells without modifying the configuration of the system. The imaginary penetration depth $\lambda_I << \lambda_R$ and dissipation is very small. In this regime, the curvature of the pinning potentials or Labusch constant $\alpha_L$ may be estimated by measuring the linear real penetration depth $\lambda_R$. Moreover, in this limit, the inductive component of the ac susceptibility χ' is only determined by the experimental geometry and an adimensional parameter $\lambda_R/D$ where $D$ is the characteristic length of the sample

To evaluate $\lambda_R$ and $\alpha_L$ in the Campbell regime, we approximated our experimental geometry by a thin disk of radius $R$ and thickness $\delta$ in a transverse ac magnetic field. We used the numerical solution developed by Brandt[38,39], in which χ is determined by the adimensional parameter $\lambda_{ac}/D$, where $D = (\delta R/2)^{1/2}$. In the Campbell regime $\lambda_R = (\lambda_L{}^2 + \lambda_c{}^2)^{1/2}$, where $\lambda_L$ and $\lambda_c = (\phi_0 B/4\pi\alpha_L)^{1/2}$ are the London and Campbell penetration depths respectively and $\phi_0$ and $B$ are the flux quantum and the magnetic induction[37]. When the phase $\varepsilon = \lambda_I/\lambda_R << 1$, to first order in $\varepsilon$, we obtain

$$\chi' + i\,\chi" = -1 + f(\lambda_R/D) + i\,\varepsilon\,g(\lambda_R/D) \quad (1)$$

where $f$ and $g$ are functions of the adimensional variable $\lambda_R/D$[38,39]. Therefore, in this limit, the inductive component of the ac susceptibility χ' is determined by the experimental geometry and the adimensional parameter $\lambda_R/D$, and we can obtain $\lambda_R$ by inverting the real part of Eq. 1:

$$\lambda_R/D = f^{-1}(\chi' + 1) \quad (2)$$

and considering $B \sim H$

$$\alpha_L = \frac{\phi_0 H}{4\pi(\lambda_R{}^2 - \lambda_L{}^2)} \quad (3)$$

The normalized London penetration depth $\lambda_L/D$ was estimated by extrapolating to $H = 0$ the $(\lambda_R/D)^2(H)$ curves (for further numerical details, see Ref.[39]).

## III. RESULTS AND DISCUSSION

The general behavior of the peak effect varying the applied ac field is depicted in Figure 1, where curves of χ"(T) (Figure 1a) and χ'(T) (Figure 1b) are shown for various $h_a$. All curves were measured at $f = 30\,kHz$ in a warming process, from a VLC prepared by cooling the sample in a dc field $H_{dc} = 2000\,Oe$ without any applied ac field ($ZF_{ac}CW$). For high amplitudes a clear PE with a minimum in χ' at the temperature $T_p$ is displayed. Below $T_p$, for the lower amplitudes, a linear response holds with a very small dissipation. As can be observed, no PE is present in this regime.

In the case of flat platelet samples in perpendicular ac fields, we should first elucidate whether geometrical-barrier (GB) dominated VL response[25] or bulk pinning dominated VL response describes our results. In the case of geometrical-barrier (GB) VL response[25], for a given $h_a$, a maximum in the non-linear χ" is expected at a dc field $H_0 \sim H_p^2(T)/h_a$, where $H_p$ is the first flux entry penetration magnetic field. At $H_{dc} << H_0$, a linear reversible response arising from GB's should hold.

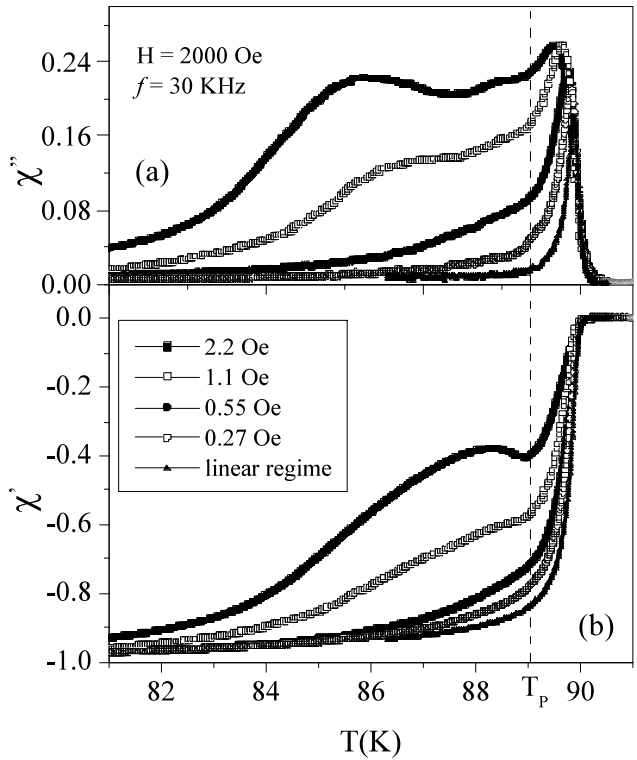

FIG. 1: Ac susceptibility components χ"(T) in (a) and χ'(T) in (b) in a $ZF_{ac}CW$ warming process (see text) for different ac measuring fields at $f = 30\,kHz$ and $H_{dc} = 2000\,Oe$. The PE with a minimum in χ'(T) at $T_p$ is well established for the non-linear response at high ac measuring field, while it is absent in the linear regime at low ac field.

The following arguments discard this option in the present situation. The calculated temperature where $H_0 \sim H_{dc} = 2000\,Oe$ for an ac field $h_a = 2.2\,Oe$, is $T \sim 26K$, well below the temperature where the peak in χ" is observed in Figure 1 (curve with black squares). Additionally, the $H_{dc}$ fields where the χ" peaks are actually observed at different temperatures, are around 100 times higher than the calculated $H_0(T)$. Consistently, in our past work, memory effects in the non-linear ac regime in twinned YBCO crystals have been successfully described within a bulk critical state model[29,30]. Therefore, we discard GB's as a possible description for the observed non-linear response.

Let us now analyze the linear response. Calculated $H_0(T)$ for the smallest ac field, $h_a = 40m\,Oe$ (linear regime), results to be smaller than the applied field $H_{dc} = 2000\,Oe$ in the whole temperature range shown in Figure 1. The observed linear response therefore holds for $H_{dc} > H_0$ and then cannot be ascribed to GB's. We therefore consistently assume bulk pinning in what follows.

As was pointed out in the introduction, above the on-set of the PE a more disordered (and more pinned) VL is expected. Changes in the number and distribution of VL dislocations may affect both static and dynamic properties. However, results from Figure 1 indicate that the pinning static properties are not so strongly affected as to produce a PE anomaly in the critical current density, even though the VL mobility is drastically modified.

A first argument for the absence of the PE in the linear response is the fact that changes in the VL topology could be related to very small changes in the Labusch constant, impossible to be detected with our experimental technique. We rule out this possibility: In Refs.[32] and[33], we have explored the pinning potential corresponding to different VLCs resulting after various protocols or dynamical histories. A clear difference in the penetration depth of the various VLCs has been observed. Therefore, we have shown that we are able to measure changes in $\lambda_R$ (i.e. in $\alpha_L$) that can only have their origin in a different VL topology (i.e. a change in the number or distribution of dislocations), and we have the sensitivity to measure that difference.

A possible second explanation for the observed behavior could be the nature of the Campbell regime. In this regime, vortices remain oscillating around a pinning position, and can only move in a very slow thermal relaxation process[33]. Vortices cannot perform large excursions and therefore most of the VLCs are not accessible. To test this possible scenario, we explored the linear response allowing vortices to alternatively move out of their pinning sites. To achieve this purpose, we performed simultaneous measurements in the linear and non-linear regime alternatively applying high and low measuring ac field amplitudes. With the higher amplitude (non-linear measurement) vortices move and may reorganize in different VLCs. With the small amplitude (linear measurement), the Labush constant corresponding to the attained configuration is measured.

In Figure 2, curves for $\chi''(T)$ and $\chi'(T)$ in a unique $ZF_{ac}CW$ warming process at $H_{dc} = 2200$ $Oe$, switching alternatively from high 1.2 $Oe$ (non-linear) to low 44 $mOe$ (linear) measuring $h_a$ amplitudes at 30 $kHz$, are shown. Even if vortices move out of their pinning sites (and explore different VLCs) when forced by a larger ac excitation that drives the non-linear response, the temperature dependence of the linear $\chi'$ (right axis in the figure) remains monotonous.

Motivated by this striking fact, we have performed an exhaustive comparison of the linear and non-linear behavior in the PE region. In Figure 3, non-linear (top panel) and linear (lower panels) response as a function of temperature at various $dc$ fields are compared. Figure 3a shows non-linear $\chi'(T)$ for various $ZF_{ac}CW$ warming processes. In all the cases, the PE is present. In Figure 3b the calculated $\alpha_L(T, H_{dc})$ is plotted in logarithmic scale as a function of temperature for the different $dc$ fields. No dramatic changes are observed in the $T$ dependence of the pinning potential curvature, $\alpha_L$, for any $dc$ field. In Figure 3c the linear $\chi''(T)$ is displayed. By inspecting the figure however, two characteristic temperatures are observed. At a first temperature we denominate $T_1(H_{dc})$ (see upward arrows in panel b) there is a subtle modification in the curvature of $\alpha_L(T)$ at fixed field, that leads to a change in the dependence of $\alpha_L(H)$ at fixed temperature. Below this temperature, $\alpha_L$ seems to become field independent for magnetic fields $H \leq H_{dc}$ within our experimental resolution whereas above $T_1(H_{dc})$, $\alpha_L$ grows with $H$ for $H \geq H_{dc}$. Upward arrows in the figure indicate $T_1$ for the lower fields: $T_1(500\ Oe) \sim 87.75$ $K$ ( full gray arrow) and $T_1(250\ Oe) \sim 89.1$ $K$ ( full black arrow). We are not able to measure $\alpha_L(T)$ at low temperatures because $\lambda_C$ becomes too small, so that the corresponding $T_1(H_{dc})$ for the higher fields lies outside to the left of the measured range. At constant $H_{dc}$, because all the superconducting parameters involved in pinning decrease, the Labusch parameter decreases smoothly and monotonically with temperature in the whole temperature range, but at $T_1(H_{dc})$ the decrease becomes slower, producing a qualitative change in $\alpha_L(H)$ in the PE region.

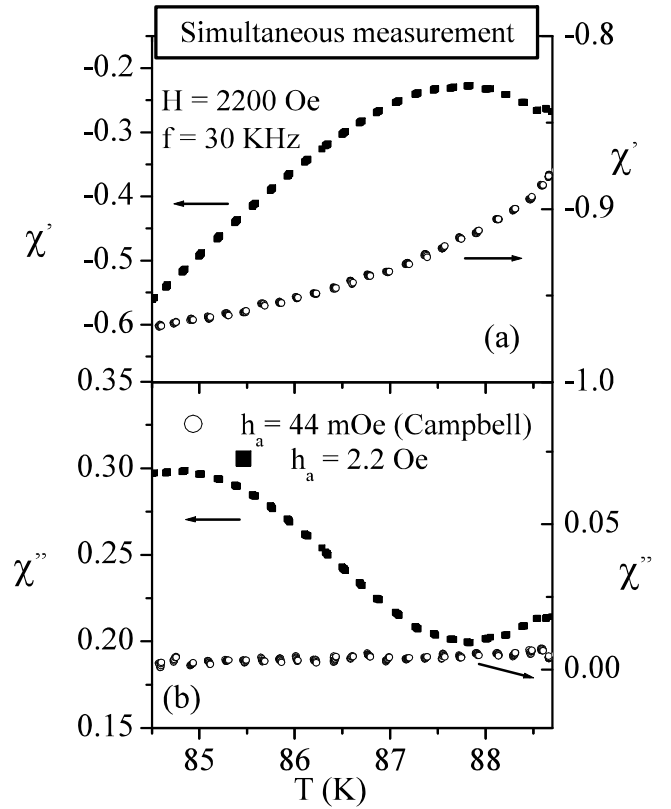

FIG. 2: $\chi'(T)$ and $\chi''(T)$ in a unique $ZF_{ac}CW$ warming process for $H_{dc} = 2200$ $Oe$, switching alternatively from high ($h_a = 1.2$ $Oe$, full squares, left axis) to low ($h_a = 44$ $mOe$, open circles, right axis) measuring field amplitudes, at $f = 30$ $kHz$. Although intervalley vortex excursions in the non-linear regime (where the PE is observed) allow to explore different VL configurations, the PE is absent in the linear regime.

At a higher temperature $T_2(H_{dc})$ there is a sudden increase in the linear $\chi''$ shown in Figure 3c (see downward arrows), indicating the end of the Campbell regime and the onset of dissipative mechanisms. We fixed $\chi'' = 0.014$ that corresponds approximately to a phase $\varepsilon \sim 0.07$ as a criterion to define $T_2$. Downward arrows in Figure 3c show $T_2(2000\ Oe) = 89.1\ K$ (dashed arrows), $T_2(1500\ Oe) = 89.55\ K$ (full gray arrows) and $T_2(1000\ Oe) = 89.8\ K$ (full black arrows). Note that $T_2(H_{dc})$ is just above the minimum in the non-linear $\chi'(T)$ shown in the top panel, and coincides with the end of the PE.

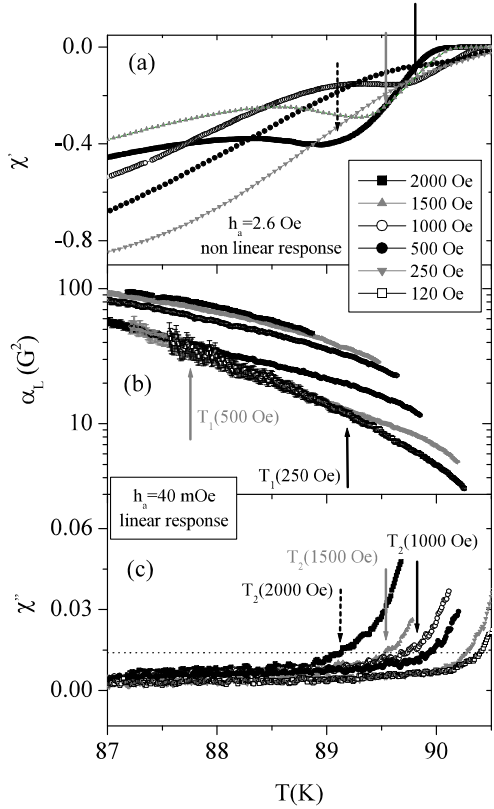

FIG. 3: (a) Non-linear $\chi'(T)$, showing the PE for different $H_{dc}$. (b) $\alpha_L(T)$ in logarithmic scale for different $H_{dc}$. (c) Linear $\chi''(T)$ indicating the Campbell regime range ($\chi''(T) \sim 0$). Upward arrows in panel (b) indicate $T_1(H_{dc})$ ($T_1(500\ Oe)$ gray and $T_1(250\ Oe)$ black arrows), where $\alpha_L$ starts to increase with $H$ (see text). Downward arrows in panels (a) and (c) show $T_2(H_{dc})$ ($T_2(2000\ Oe)$ dashed arrows, $T_2(1500\ Oe)$ full gray arrows and $T_2(1000\ Oe)$ full black arrows), where the linear $\chi''(T)$ starts to grow (see text).

To better examine the field dependence, data in Figure 3 at a few selected temperatures, have been plotted in Figure 4 as a function of the applied $dc$ magnetic field. The comparison of linear ( lower panels) and non-linear response (top panel) is shown. In Figure 4a the non-linear $\chi'(H)$ with a broad PE in dc field for the highest temperatures is shown. In Figure 4b values of $\alpha_L(H)$ are shown in logaritmic scale. In figure 4c the linear $\chi''(H)$ is displayed. Note that the highest temperatures were selected in way to match with $T_1$ or $T_2$ for some of the measured fields (e.g. 87.75 K ~ $T_1(500 Oe)$, 89.1K ~ $T_1(250\ Oe) = T_2(2000\ Oe)$, 89.55 K = $T_2(1500\ Oe)$, 89.8 K = $T_2(1000\ Oe)$).

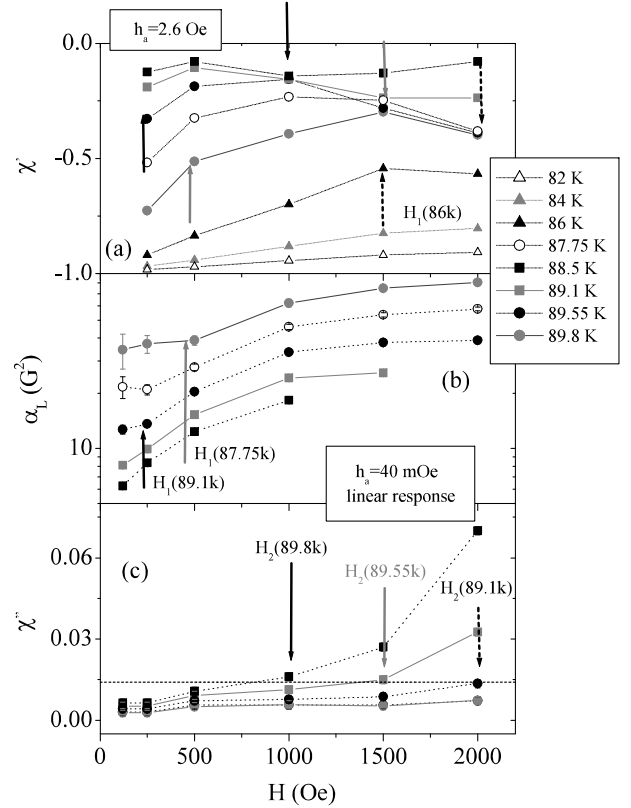

FIG. 4: (a) Non-linear $\chi'(H)$ showing a broad PE for different selected temperatures. (b) $\alpha_L(H)$ and error bars (see text) for different $T$. in logarithmic scale. (c) Linear $\chi''(H)$ indicating the Campbell regime range ($\chi''(H) \sim 0$). Upward arrows in panel (b) are the same upward arrows shown earlier in Fig. 3 indicating $H_1(T)$ ($H_1(89.1K) \sim 250\ Oe$, $H_1(87.75\ K) \sim 500\ Oe$ full gray and black arrows), where $\alpha_L(H)$ begins to increase strongly with field, setting the onset of the PE in the non-linear $\chi'(H)$. $H_1(86K) \sim 500\ Oe$ is estimated (see text). Downward arrows in panels (a) and (c) are the same downward arrows of Fig. 3. They indicate $H_2(T)$ ($H_2(89.8K) \sim 1000\ Oe$ full black, $H_2(89.55\ K) \sim 1500\ Oe$ full gray and $H_2(89.1K) \sim 2000\ Oe$ dashed arrows), where Campbell regime is lost and the minimum in the non-linear $\chi'(H)$ occurs. In all cases lines are a guide to the eye.

The same arrows from Fig. 3 that indicated $T_1(H_{dc})$ and $T_2(H_{dc})$, are shown in Figure 4, indicating now $H_1(T)$ and $H_2(T)$. We identify then two characteristic magnetic fields: At a first magnetic field $H_1(T)$ we find a sudden change in the dependence of $\alpha_L(H)$, that begins to grow with $H$, as was anticipated in figure 3b. Coincidently, the rate of increase of the non-linear $\chi'(H)$ begins to be reduced. The last fact implies a reduction in the rate of increase in the VL mobility and/or a reduc-tion in the rate of decrease of the measured frequency dependent $J_c(H,T,f)$, producing the onset of the PE. Upward arrows in Figure 4 indicate: $H_1(89.1\ K) \sim 250\ Oe$ (full black arrows) and $H_1(87.75\ K) \sim 500\ Oe$ (full gray arrows). By inspecting figure 4a, we find an additional $H_1(T)$ in the range of low temperatures, where $\alpha_L$ is not measurable, by identifying approximately the onset of the PE. The dashed gray upward arrow shows $H_1(86\ K) \sim 1500\ Oe$.

To analyze the upper bound of the PE, we refer now to panel c) in figure 4, where at $H_2(T)$ the linear $\chi"$ grows up to the limit criterion and Campbell regime is lost. Coincidently, the non-linear $\chi`(H)$ in figure 4a displays a broad minimum identifying the end of the PE. Downward arrows in the figure show : $H_2(89.8\ K) \sim 1000\ Oe$ (full black arrows), $H_2(89.55\ K) \sim 1500\ Oe$ (full gray arrows) and $H_2(89.1\ K) \sim 2000\ Oe$ (dashed arrows). Therefore, the PE region where the anomaly in the non-linear response develops, occurs between the fields $H_1$ and $H_2$. The growth of $\alpha_L$ with $H$ above $H_1$ may be related to a reduction of the correlation volume due to the increase of the number of VL defects in the PE region. The loss of the Campbell regime at $H_2$, indicates that above this field, dissipative forces prevail over pinning forces in the linear regime.

The proposed picture to explain the above results is the following: For $H > H_1(T)$ the VL softens and there is a continuous increase in the number of VL defects producing a decrease in the correlation volume and an increase in the Labusch constant with $dc$ field. This mechanism competes with the intrinsic decrease of $\alpha_L$ with temperature, but is not strong enough to overcome it. Therefore, the anomaly observed in the non-linear penetration depth with temperature, cannot be ascribed only to the softening of the VL, allowing a better accommodation to the pinning centers. The increase in VL defects produces a drastic change in the dynamics and/or creep mechanism, and gives rise to the observed PE anomaly in the non-linear response.

As was already pointed out[16,17], a dynamic origin for the PE should be related to time dependent phenomena. In a continuum description, the dynamics of the VL is directly related with the constitutive relation between the electric field $E$ and the local time dependent current density $J$. In an $ac$ susceptibility experiment $h_a << H$ and $B \sim H$, then the constitutive relation $E(J)$ is expected to depend only on the applied magnetic field and temperature, that together with the sample geometry determines the $\chi"(\chi`)$ curves. However, the increasing evidence of history effects in VL dynamics should lead, within this framework, to include history effects in parameters contained in the constitutive relation.

Well known microscopic models for vortex dynamics[7] yield a current density dependent activation energy given by $U(J) = U_0[(J_c/J)^\mu - 1]/\mu$, with $\mu > 0$, $J_c(B,T)$ is the critical current density and $U_0(B,T)$ is the energy scale for the barriers. For thermal creep, $E(J) = E_c \exp(-U(J)/kT)$, where k is the Boltzmann constant. For small values of $\mu$, a suitable approximation for the activation energy is[7] $U(J) \approx U_0 ln(Jc/J)$, that leads to the well known power law dependence constitutive relation:

$$E(J) = E_c(J/J_c)^n \qquad (4)$$

The exponent $n = U_0/kT$, is field and temperature dependent, and the critical current density $J_c$ is defined with the voltage criterium $E_c$. Theoretical curves for $\chi"(\chi`)$ were calculated in Ref.[38] for a type II superconducting thin disk with the power law constitutive relation given above. In this model, $\chi"(\chi`)$ is predicted to depend only on the exponent $n$ and the non-linear $ac$ penetration depth. If Eq. 4 holds, experimental points are expected to lie in a curve that corresponds to a given $n$. The limit $n = 1$ corresponds to the Ohmic Flux Flow (FF) response, and $n \to \infty$ to an ideal Bean critical model. In a typical $ac$ susceptibility experiment as a function of temperature, the $\chi"(\chi`)$ curve is not a constant $n$ curve. It will correspond to $n \approx 1$ at high temperatures near the FF regime and to $n \gg 1$ at low temperatures, near the critical state response. However, measurements at fixed $T$ and $H$ and at different frequencies should lie in a constant $n$ curve. We have then performed measurements at several frequencies that are discussed below.

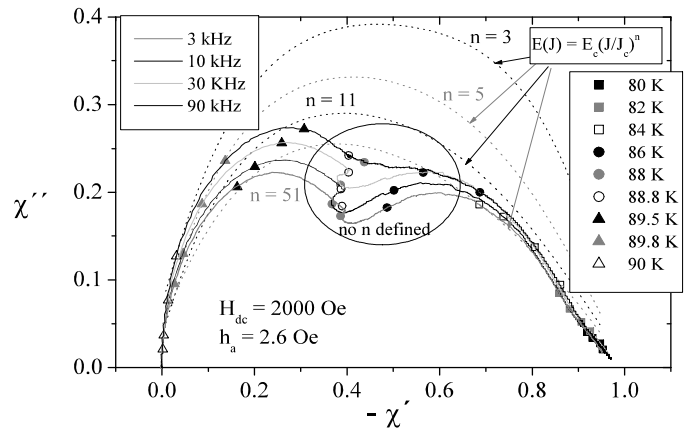

FIG. 5: $\chi"(\chi`)$ plot. Dotted lines are calculations of $ac$ susceptibility for a disk including creep, for the constitutive relation $E(J) = E_0(J/J_c)^n$ describing logarithmic current dependent creep barriers. In full lines are the curves for different frequencies $f = 3, 10, 30$ and $90\ kHz$ for $H_{dc} = 2000\ Oe$ and $h_a = 2.2\ Oe$. Some fixed temperatures are indicated in symbols. Low $T$ data are close to a Bean critical state, high $T$ data are close to flux flow regime. Note that in the encircled PE region, no $n$ describes VL dynamics.

Figure 5 was constructed plotting $\chi"(\chi`)$ from experimental non-linear curves $\chi"(T)$ and $\chi`(T)$ for $f = 3, 10, 30$ and $90 KHz$. All curves were measured for $H_{dc} = 2000\ Oe$ and $h_a = 2.6\ Oe$ (solid lines in the figure).

Within these curves, points corresponding to various selected temperatures are plotted with a single symbol for each $T$. Experimental data are compared with curves predicted by Eq. 4 for several $n$ (dashed lines in the figure).

It can be seen that at low temperatures, $\chi"(\chi')$ curves approach a Bean model with an effective $J_c(f)$[7]. It is evident that for the $85\ K \lesssim T \lesssim 89\ K$ data (encircled region in the figure) $n$ is not defined; experimental curves are qualitatively different from any theoretical curve calculated using Eq.(4), indicating an abrupt change in VL dynamics. However, for $T > 89.4\ K$ (triangles in the figure) a valid $n$ is recovered that decreases rapidly with temperature towards an ohmic regime, with $n = 1$ in the unpinned vortex liquid.

To compare the temperature range between 85 and 89 $K$, with $T_1(H = 2000\ Oe)$ and $T_2(H = 2000\ Oe)$ we go back to figures 3 and 4. Note from Figure 3 that $T_2(H_{dc} = 2000\ Oe) = 89.1\ K$. On the other hand, in Figure 4 we have estimated $H_1(86\ K) \sim 1500\ Oe$ whereas $H_1(84\ K)$ is not observed for $H < 2000 Oe$, implying that $84\ K < T_1(H = 2000\ Oe) < 86\ K$. Therefore, the region where there is an abrupt change in the VL dynamics lies between $T_1(H)$ and $T_2(H)$, that is the region of the Peak Effect.

This result closes the proposed picture. In YBCO crystals, the PE as a function of $H$ occurs between two characteristic fields $H_1(T)$ and $H_2(T)$ due to a drastic change in the VL dynamics. This fact coincides with a clear variation in the $H$ dependence of $\alpha_L$, probably related to a continuos increase in the number of dislocations. In this description, the observed change in the dynamics may be due to a transition from elastic to plastic creep with a larger $\mu$[4], where the relation 4 would be no longer valid.

## IV. CONCLUSIONS

We have explored the static and dynamic behavior of the vortex lattice in YBCO crystals, in the PE region, at low fields and high temperatures. By performing sensitive linear and non-linear $ac$ susceptibility measurements we have compared the behavior of the measured critical current and the effective pinning potential wells as a function of temperature, magnetic field and frequency.

For the lower amplitudes, a frequency independent linear Campbell response holds with a very small dissipation. The PE (non- monotonous $T$ dependence) appears only in the non linear curves with high dissipation, where vortices perform inter-valley motion and an important influence of thermal creep in the response is expected, whereas in the linear Campbell regime no PE is observed.

We conclude that the PE as a function of magnetic field occurs between two characteristic fields $H_1(T)$ and $H_2(T)$. At the first field $H_1$, the Labusch constant $\alpha_L$ shows an increase with field. Below $H_1$, $\alpha_L$ seems to be field independent within our resolution. The increase of $\alpha_L$ with $H$ above $H_1$ may be related to a reduction of the correlation volume due to the increase in the number of VL defects. The sudden change in $\alpha_L(H)$ coincides approximately with the beginning of the anomaly in the slope of the non-linear $\chi'(H)$ curves.

At the second field $H_2$, the linear real penetration depth grows faster and the linear dissipative component suddenly increases, indicating that dissipative forces prevail over the pinning forces and the Campbell regime is lost. The loss of the Campbell regime coincides with the minimum observed in the non-linear $\chi'(H)$ curves, indicating the end of the PE region. We interpret this as a signature of the transition to a vortex liquid.

As a function of temperature, at constant magnetic field, we identify two characteristic temperatures $T_1(H_{dc})$ and $T_2(H_{dc})$ as those points where the applied $H_{dc}$ coincides with $H_1$ and $H_2$ respectively. Above $T_1(H_{dc})$ there is a continous reduction of the correlation volume that competes with the intrinsic decrease of $\alpha_L$ with temperature, but is not enough to overcome it. We emphasize that this fact indicates that the PE cannot be ascribed only to an increase in the pining force (i.e. a softening to the VL, allowing a better accommodation to the pinning centers). An additional process must cooperate to enhance the anomaly in the non-linear regime.

We have compared experimental non-linear $\chi"(\chi')$ curves at constant $H$ with theoretical curves predicted for a thin disk in the presence of creep, with a power constitutive relation $E(J)$. The data were extracted from measurements at various frequencies. At low temperatures $\chi"(\chi')$ curves follow a Bean dependence, and above $T_2$ a transition to an Ohmic regime (characterisic from an unpinned vortez liquid), passing through a highly non-linear dissipative intermediate regime occurs.

However, for $T_1 < T < T_2$ experimental curves are qualitatively different from the predicted theoretical curves indicating that an abrupt change in VL dynamics occurs in the PE region. We find that the dynamics in this region can not be described within the model, probably due to the relevance of plastic creep.

The smooth behavior in $\alpha_L(T)$ indicates that there is not a sudden change in the correlation volume at the onset of the PE, as would be expected in a first order-disorder phase transition at that point. Furthermore, the abrupt increase of dissipative mechanisms at $T_2$, suggests a transition at the end of the PE. However, we point out that the existence and nature of a phase transition cannot be provided by $ac$ susceptibility measurements, and this issue deserves future studies.

Summarizing, we propose that the Peak Effect in YBCO crystals, arises from a drastic change in the dynamics between two characteristic H-T lines, $H_1(T)$ and $H_2(T)$. This change is probably due to a continuos increase in the number of dislocations in that region, favored by a softened VL, before the transition to a vortex liquid occurs.